\documentclass[reprint,
showpacs,
 amsmath,amssymb,
 aps,
 prl,
]{revtex4-1}

\usepackage{amssymb}
\usepackage{amsthm}
\usepackage{graphicx}
\usepackage{dcolumn}
\usepackage{bm}
\usepackage{amsfonts}
\usepackage{amsmath}
\usepackage{amssymb}
\usepackage{natbib}
\usepackage{verbatim}

\usepackage[dvipsnames]{xcolor}

\begin{document}


 \title{Spontaneous chiralization of polar active colloids}
 
 
 \author{Marco De Corato}
\email{mdecorato@unizar.es}
\affiliation{Aragon Institute of Engineering Research (I3A), University of Zaragoza, Zaragoza, Spain}


\author{Ignacio Pagonabarraga}
\email{ipagonabarraga@ub.edu}
\affiliation{Departament de F\'{i}sica de la Mat\`{e}ria Condensada, Universitat de Barcelona, C. Mart\'{i} Franqu\`{e}s 1, 08028 Barcelona, Spain \\
University of Barcelona Institute of Complex Systems (UBICS), Universitat de Barcelona, 08028 Barcelona, Spain \\
CECAM, Centre Européen de Calcul Atomique et Moléculaire, École Polytechnique Fédérale de Lasuanne (EPFL), Batochime, Avenue Forel 2,1015 Lausanne, Switzerland\looseness=-1}

\author{Giovanniantonio Natale}
\email{gnatale@ucalgary.ca}
\affiliation{Department of Chemical and Petroleum Engineering, University of Calgary, 2500 University Drive NW, Calgary, Canada}

\begin{abstract}


Polar active particles constitute a wide class of synthetic colloids that are able to propel along a preferential direction, given by their polar axis. Here, we demonstrate a generic self-phoretic mechanism that leads to their spontaneous chiralization through a symmetry breaking instability. We find that the transition of an active particle from a polar to a chiral symmetry is characterized by the emergence of active rotation and of circular trajectories. We show that the instability is driven by the advection of a solute that interacts differently with the two portions of the particle surface and it occurs through a supercritical pitchfork bifurcation. 




\end{abstract}

\maketitle

The development of engineered active colloids that harness the chemical energy of the environment to move \cite{needleman2017active, zottl2016emergent, illien2017fuelled} has enabled to mimic and dissect mechanisms in biological systems while opening doors to multiple applications: bioremediation, micro-mixing, micro-machinery, drug delivery and more \cite{ das2015boundaries, maggi2016self,huang2018bacteria,parmar2018micro,hortelao2018targeting,wang2020biomimicry,tang2020enzyme}.
To imitate the intrinsic asymmetry of flagellates and other microorganisms, active colloids are designed with fore-aft asymmetric chemical activity, which define their polar axis and their preferential direction of motion \cite{paxton2004catalytic,howse2007self}. Alternatively, in the absence of a built-in asymmetry, a polar axis defining the direction of motion of an isotropic active particle can emerge from a spontaneous symmetry-breaking instability \cite{de2013self, michelin2013spontaneous,izri2014self,maass2016swimming,boniface2019self,hu2019chaotic,PhysRevFluids.5.122001} that is reminiscent of that used by cells to migrate \cite{RUPRECHT2015673,PhysRevLett.116.028102,farutin2019crawling}. However, unlike their biological counterparts that have evolved internal mechanochemical processes to actively change the direction of motion \cite{son2013bacteria}, active colloids can only rely on passive rotational diffusion \cite{howse2007self}. Providing them with active rotation requires to break the polar symmetry through the use of external fields or by designing them with a chiral shape or with a chiral chemical activity \cite{kraft2013brownian, kummel2013circular,wykes2016dynamic,aubret2017eppur,brooks2018shape,aubret2018diffusiophoretic,lisicki2018autophoretic,brooks2019shape,lee2019directed,reigh2020active}. The resulting self-rotation has been exploited to develop autonomous sorting of active particles \cite{mijalkov2013sorting,su2019disordered,levis2019activity,PhysRevLett.125.238003} and also as a potential solution to move barriers or micro machinery \cite{liao2018transport,aubret2018targeted}.

Recent experiments showed that Janus active particles can spontaneously break their polar symmetry and transition from a persistent Brownian motion with enhanced rotational diffusion \cite{gomez2016dynamics,aragones2018diffusion,lozano2019active,qi2020enhanced,theeyancheri2020translational,PhysRevLett.125.258002} to chiral trajectories \cite{narinder2018memory} attributed to the viscoelastic  response of the  medium in which colloids propel. Such symmetry breaking instability could offer a controllable way to trigger active rotation, which does not require a built-in chiral shape nor a chiral chemical activity. 

In this letter, we demonstrate that chiral motion is a generic feature of polar active colloids that propel through self-phoresis \cite{moran2017phoretic} and arises spontaneously from a symmetry breaking instability.  
Therefore, our findings are relevant to most of the Janus active colloids used in the experiments.



\begin{figure}[h!]
\centering
\includegraphics[width=0.45\textwidth]{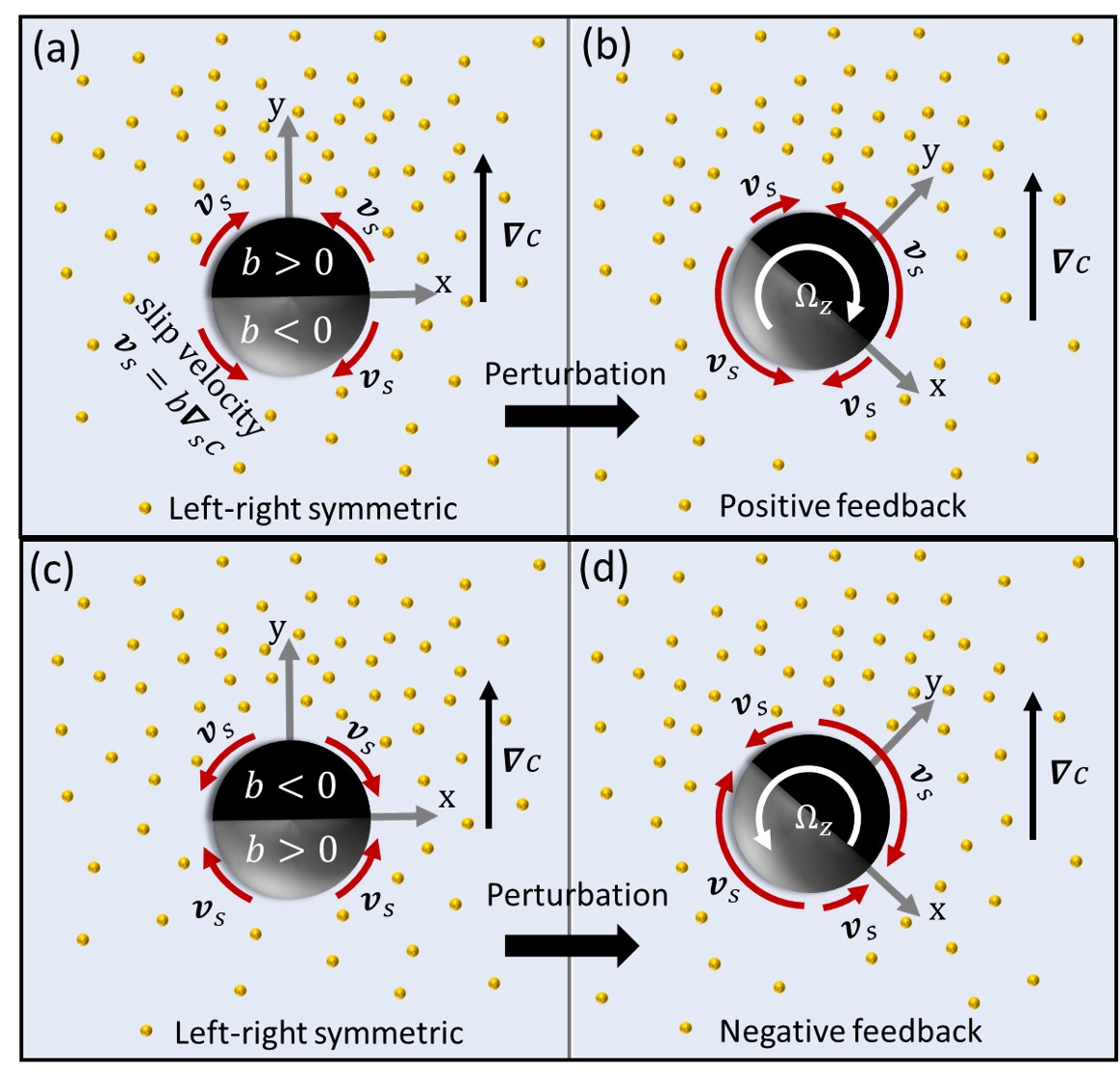}
\caption{Schematics of the spontaneous chiralization mechanism, the translational velocity is not shown for clarity. (a) and (c): a Janus particle with positive and negative phoretic mobility, $b$, on the two hemispheres and a larger chemical activity on the black hemisphere is in a left-right symmetric configuration. $b>0$ and $b<0$ signify repulsive and attractive solute-wall interactions, respectively. (b) and (d): a small perturbation of the particle polar axis leads to configurations in which the active particle has a net rotation rate. To conserve the angular momentum, the hydrodynamic torque generated by the slip velocity is balanced by an opposite rotation of the particle. The direction of the hydrodynamic torque is determined by the direction of the slip flow $\boldsymbol{v}_s$. In (b), the slip flows generate a counterclockwise torque leading to a clockwise particle rotation, which reinforces the initial perturbation possibly breaking the left-right symmetry. In (d), the phoretic mobility, $b$, is flipped between the two hemispheres, the slip flow is reversed and the particle rotation restores the initial left-right symmetric configuration. }
\label{fig1}
\end{figure}

To illustrate the mechanism leading to spontaneous chiralization, we consider a spherical colloidal particle of radius $R$ that is suspended in a Newtonian and incompressible liquid of shear viscosity $\eta$. We employ a co-rotating Cartesian reference frame with origin in the particle center. Without any loss of generality we choose the $y$ axis to be aligned with the polar axis of the active particle (See Fig. \ref{fig1}). The particle is chemically active and releases or consumes a solute specie at its surface.
The solute surface flux density, of magnitude $Q$, is characterized by its axisymmetric surface distribution $f(y/R)$. 

\begin{figure*}[t]
\centering
\includegraphics[width=1.0\textwidth]{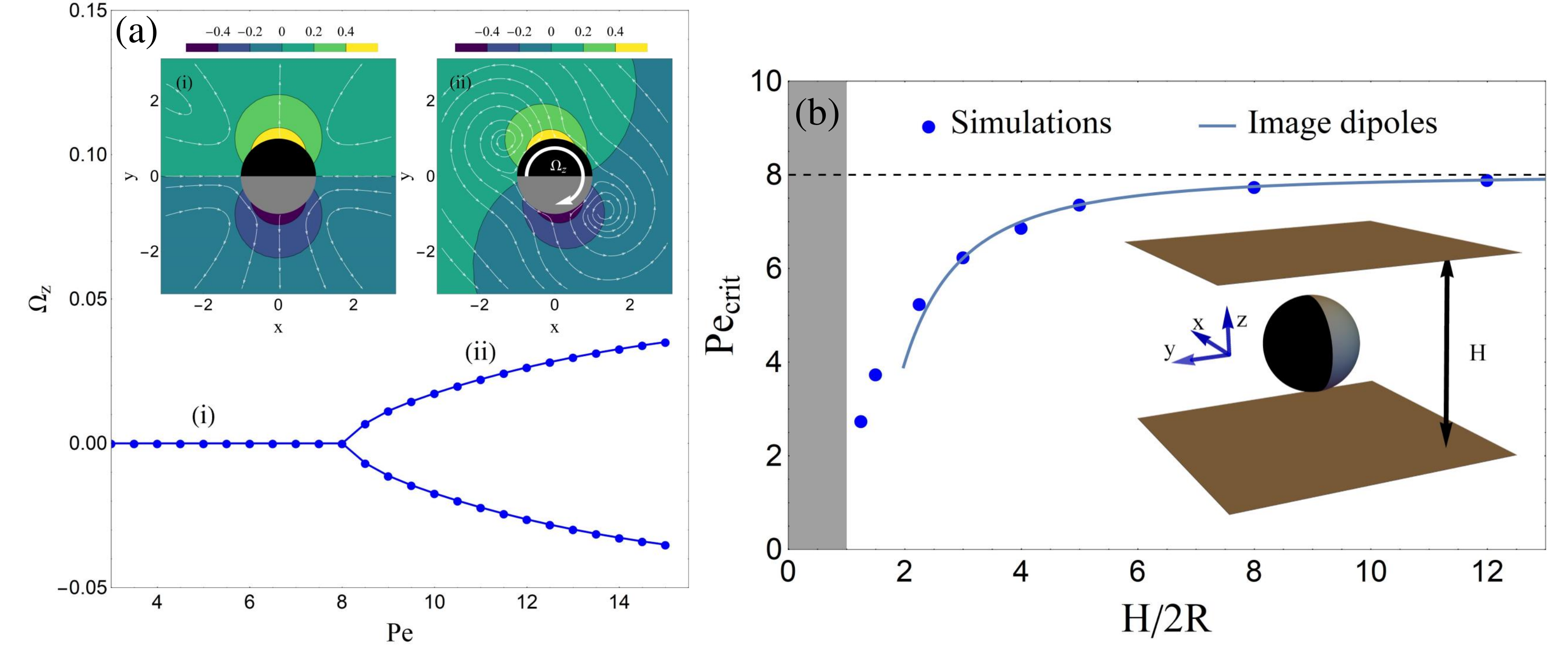}
\caption{(a) The rotational velocity made dimensionless using the characteristic time $Qb/RD$, plotted as a function of the P\'{e}clet number in the case of $f(y/R)=g(y/R)=y/R$. For $Pe<8$ the concentration profile is left-right symmetric and aligned with the polar axis of the particle. For $Pe>8$ the polar symmetry breaks and the particle starts to rotate. Insets: distribution of solute in the polar (i) and in the chiral regime (ii). (b) $Pe_{crit}$ is strongly reduced by confining the particle between two parallel walls (see text).}
\label{fig2}
\end{figure*}

To investigate the robustness of the symmetry breaking mechanism, we consider both that a self-diffusiohoretic particle displaces freely in an unbounded  liquid and confined in the symmetry plane, $z=0$, between two parallel walls located at $z=\pm H/2$, see the inset of the right panel of Fig. \ref{fig2}.  
In the co-moving reference frame, the steady state solute concentration, $c$, obeys:
\begin{equation} \label{transp_1}
 D \boldsymbol{\nabla}^2 c  = \boldsymbol{v} \cdot \boldsymbol{\nabla}c \, \, ,
\end{equation}
with $\boldsymbol{v}$ the fluid velocity and $D$ the solute diffusion coefficient.
Far from the particle the concentration of the solute is kept at $c=0$ and the walls are impermeable to the solute. The flux at the surface of the particle, $r=R$, is given by $\boldsymbol{n} \cdot \boldsymbol{\nabla}c  =-Q \, f(y/R)/D$, with $\boldsymbol{n}$ the vector normal to the surface of the sphere pointing outwards. On the portions of the surface where $f(y/R)>0$, the solute is released, while the solute is consumed if $f(y/R)<0$.

The colloid displaces through self-diffusiophoresis, determined by the phoretic slip velocity, $\boldsymbol{v}_s= b \, g(y/R) \boldsymbol{\nabla}_s c$, with magnitude $b$, and where $\boldsymbol{\nabla}_s$ stands for a gradient parallel to the colloid surface. This is an effective description of the flow  generated by molecular solute-surface interactions in a thin boundary layer (typically smaller than a few nanometers) around the particle surface \cite{anderson1989colloid,moran2017phoretic}. The function $g(y/R)$ encodes the effect that solute-wall interactions has on induced fluid flows, and can change both its magnitude, $b$, and sign, the latter characterizing whether solute is attracted/repelled ($g<0$/$g>0$) to the colloid, along the particle surface. However, a more general approach should be developed for ionic species \cite{de2020self} or when advection becomes important in the boundary layer \cite{michelin2014phoretic}.

The slip velocity at the boundary of the particle generates a flow field that obeys the Stokes equation:
\begin{equation}\label{mom_bal_gen}
\boldsymbol{\nabla} \cdot \boldsymbol{\sigma} =\eta \boldsymbol{\nabla}^2 \boldsymbol{v} - \boldsymbol{\nabla} p =  \boldsymbol{0}  \, \, , \, \, \boldsymbol{\nabla} \cdot \boldsymbol{v}=  0 \, \, ,
\end{equation}
with $\boldsymbol{\sigma}$ the stress tensor for a Newtonian fluid and with the boundary conditions at the surface of the particle given by $\boldsymbol{v}=  \boldsymbol{v}_s$.
In the co-moving frame, the boundary conditions at infinity and on the walls are given by $\boldsymbol{v}=  -\boldsymbol{V} - \boldsymbol{\Omega} \times \boldsymbol{r}$, where $\boldsymbol{V}$ and $ \boldsymbol{\Omega}$ are the translational and rotational velocity of the particle, respectively. 
We neglect the Brownian motion and the inertia of the active particle and we use the reciprocal theorem to compute its kinematics \cite{stone1996propulsion,masoud2019reciprocal}:
\begin{equation}\label{rectheo}
    \boldsymbol{V} \cdot \boldsymbol{F}^* + \boldsymbol{\Omega} \cdot \boldsymbol{T}^* = -\int_S \boldsymbol{v}_s \cdot \boldsymbol{\sigma}^* \cdot \boldsymbol{n} \, dS \, \, .
\end{equation}
By considering simpler auxiliary Stokes problems such as that of a sphere undergoing rigid body translation or rotation, we compute $\boldsymbol{V}$ and $ \boldsymbol{\Omega}$. In Eq. \eqref{rectheo}, the quantities $\boldsymbol{F}^*$, $\boldsymbol{T}^*$ and $\boldsymbol{\sigma}^* \cdot \boldsymbol{n}$ represent the hydrodynamic force, torque and traction acting on the sphere in the auxiliary problem. These quantities are known analytically for the translation or rotation of a freely suspended sphere \cite{leal2007advanced}, but must be computed numerically for a sphere under confinement \cite{supmat}. Finally, the integral in the right hand side of Eq. \eqref{rectheo} is carried out over the particle surface.

The problem is completely determined by the P\'{e}clet number, $Pe= Q b R/D^2$,  which compares the rate of advective phoretic flows with the rate of solute diffusion and which is based on the magnitude of the slip velocity rather than on the translational velocity of the particle \cite{michelin2014phoretic}, and by the two dimensionless functions $f\left(y/R\right)$ and $g\left(y/R\right)$, which specify the distribution of solute flux and phoretic mobility along the colloidal particle, respectively.  
In the absence of advection, $Pe=0$, Eqs.\eqref{transp_1}-\eqref{rectheo} are linear and the particle can only displace along its symmetry axis. Non-axysimmetric flows can develop at finite $Pe$. We explore numerically (details of the numerical method are provided in \cite{supmat}) the emergence of steady state rotation, which represent symmetry-broken states  where the particle spontaneously acquires chirality.

The symmetry-breaking mechanism is schematically depicted in Fig. \ref{fig1}. In Figs. \ref{fig1}(a) and (c) an active colloid is aligned with the gradient of solute generated by the chemical reaction occurring on its surface. The interaction between the solute and the surface of the particle generates a phoretic slip velocity $\boldsymbol{v}_s= b \, g(y/R) \boldsymbol{\nabla}_s c$, which cannot generate a rotation rate because the configuration is left-right symmetric. A transient perturbation of the polar vector of the particle leads to a configuration where the phoretic slip velocity applies a net torque to the fluid, which is balanced by an opposite rotation of the particle, see Figs. \ref{fig1}(b) and (d).  
If the rotation rate is directed as in Fig. \ref{fig1}(b), then the initial perturbation is reinforced. For sufficiently strong phoretic flows the perturbation grows, the polar symmetry breaks and the active particle keeps rotating, thus acquiring a chiral symmetry. Conversely, if the phoretic flows and the rotation rate of the particle are directed as in Figs. \ref{fig1}(d), the initial polar symmetry is restored.

 To illustrate how a polar active particle spontaneously acquires chirality, we investigate a simple realization of Fig. \ref{fig1} in which the chemical activity and the phoretic mobility change linearly along the polar axis of the active particle $f\left(y/R\right)=g\left(y/R\right)=y/R$, which corresponds to the case depicted in Fig \ref{fig1}(a). In this case, the solute displays repulsive interactions with the portion of surface where it is released from and it is attracted by the portion of the surface by which it is adsorbed.  
 Because of this simple choice, the active particle does not translate but generates recirculating flows \cite{golestanian2007designing}. 
 In Fig. \ref{fig2}(a), we plot $\Omega_z$ as a function of $Pe$ for a freely suspended particle. The stationary state undergoes a supercritical pitchfork bifurcation becoming unstable at a critical P\'{e}clet number. For $Pe>Pe_\text{crit}$ the particle acquires a chirality and spontaneously rotates either clockwise or anticlockwise about the z axis. The insets of Fig. \ref{fig2}(a) show the solute distribution around the particle in the polar and in the chiral steady states. In the chiral regime, the advection due to the active rotation breaks the axisymmetric distribution of solute. Such distribution generates phoretic flows that sustain the rotation of the active particle, as schematically depicted in Fig. \ref{fig1}(b). 
The spontaneous chiralization is prevented when solute is attracted to the  the fraction of the colloidal surface where it is released, $g\left(y/R\right)=-y/R$, corresponding to Figs. \ref{fig1}(c-d). Numerical analysis of this regime reveals that the stationary state is stable for all the $Pe$ considered in Fig. \ref{fig2}. For this choice of $g\left(y/R\right)$, the phoretic active rotation stabilizes the left-right symmetry against perturbations, as shown schematically in Fig. \ref{fig1}(c-d).

Fig. \ref{fig2}(b) shows that confinement promotes self-chiralization, decreasing $Pe_{crit}$, despite the increased friction  due to the walls hinders particle's rotation. As the gap between the walls is reduced, more solute accumulates around the particle, enhancing the magnitude of the self-induced phoretic flows. Since  the active particle behaves, to leading order, as a source dipole the effect of the walls  can be understood as produced by image dipoles at a distance $H/R$ beyond their actual position. The two image dipoles increase the concentration experienced by the particle  by a factor $(R/H)^2$, enhancing the induced flows by the same amount. This enhancement predicts the main decrease of $Pe_{crit}$ with fluid gap, as shown in Fig. \ref{fig2}(b). This asymptotic scaling breaks down for strong confinement because higher-order singularities are required to fulfill the boundary conditions.

\begin{figure*}[ht!]
\centering
\includegraphics[width=1.0\textwidth]{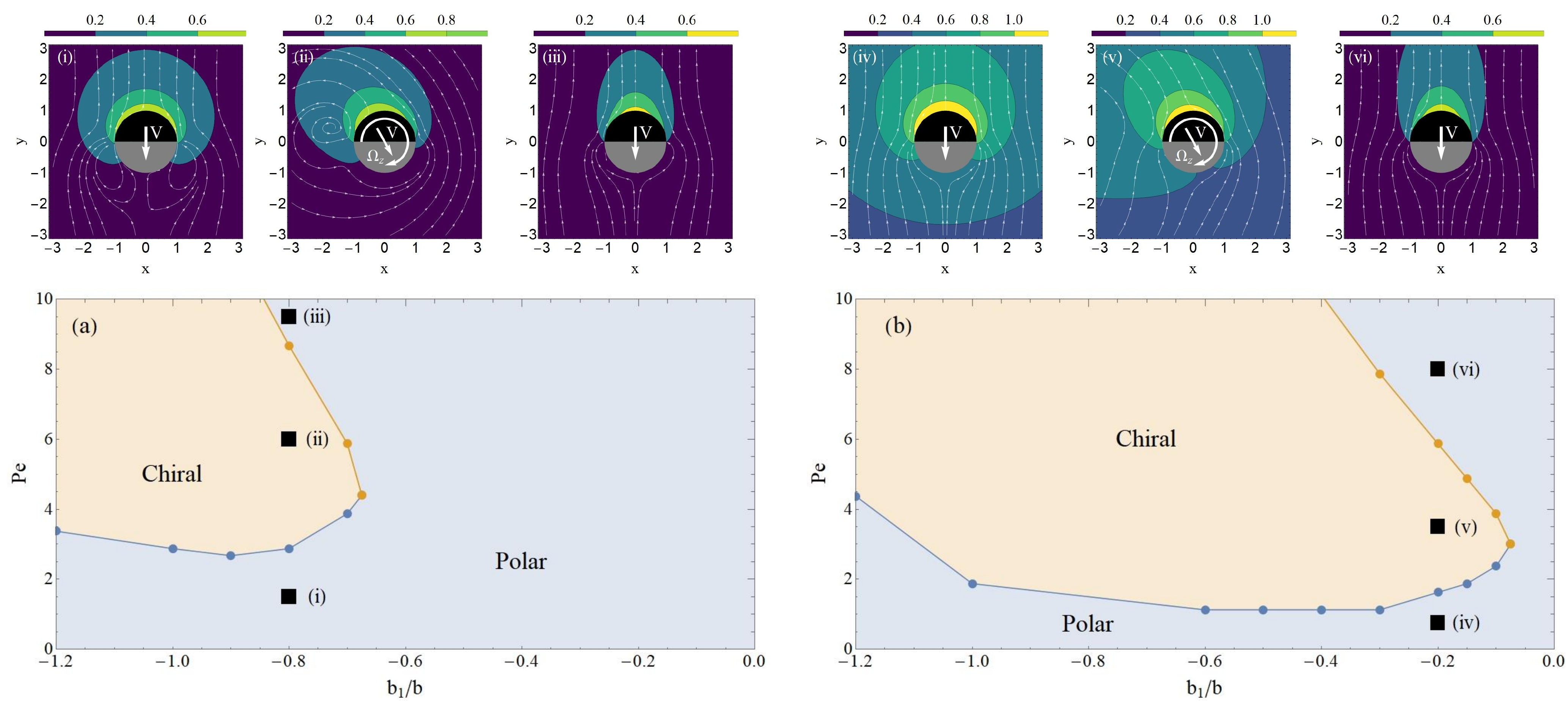}
\caption{Phase diagram of the behavior of a Janus colloid (a) freely suspended and (b) confined between two parallel walls with gap size $H=4R$. The upper figures show the steady state solute distribution and the flow streamlines in the symmetry plane $z=0$ in the case of polar symmetry (i),(iii),(iv) and (vi) and in the case of chiral symmetry (ii) and (v).}
\label{fig3}
\end{figure*} 
The apolar Janus colloid described so far highlights the essential nature of self-chiralization. However, colloids will generically self-propel when the chemical activity and phoretic mobility vary independently. We consider a Janus particle that releases solute only from one hemisphere, specifically $f(y/R)=\Theta(y/R)$ where $\Theta$ is the Heaviside theta function, and that repels solute from its catalytic side, $g(y/R)=1$ for $y>0$, and is either attractive or repulsive on its inactive side, $g(y/R)=b_1/b$; for $b_1>0$ $(b_1<0)$  the particle translates along (against) its polar axis \cite{golestanian2007designing,michelin2014phoretic}. 

Fig. \ref{fig3}(a) summarizes the behavior of  a freely suspended Janus particle as a function of $Pe$ and $b_1/b$.
The parameter space is divided in two regions: a region where the polar symmetry is stable and a region in which the polar symmetry spontaneously breaks and the active particle becomes chiral. We find that, for a broad range of $b_1/b$, exists a critical P\'{e}clet number, $Pe_\text{crit}$, above which the polar symmetry is broken.
For $Pe<Pe_\text{crit}$ and for the values of $b_1/b$ investigated, the active particle moves towards its inactive hemisphere along a straight trajectory. 
By increasing $Pe$ above its critical value, the active particle starts to rotate and its trajectory, observed from a fixed frame, abruptly changes from straight to circular. The curvature of the trajectory is proportional to  $\left|\boldsymbol{V}\right|/\Omega_z$ and, as reported in the supplementary note \cite{supmat}, is a function of $Pe$, it diverges at $Pe=Pe_{crit}$ and it is consistent with experimental observations \cite{narinder2018memory} for $Pe>Pe_{crit}$. At large $Pe$, Fig. \ref{fig3}(a) shows a re-entrant polar regime where the left-right symmetry is restored. In this regime, the motion of the particle advects the solute away from the face that is chemically inert, as shown in of Fig. \ref{fig3}(iii). In this case, most of the solute concentrates in the wake behind the active particle, hindering the phoretic flows required to sustain the active rotation and therefore stabilizing the  polar symmetry of the particle motion.  

Figure \ref{fig3}(b) displays polar and chiral regimes for a Janus particle confined in the symmetry plane between two parallel walls separated by a distance $H=4R$, which is the one used in recent experiments \cite{narinder2018memory}. The results show that the range of parameters for which the active particle displays chiral symmetry is larger compared to the case of a freely suspended particle. The confinement reduces the critical P\'{e}clet number and broadens the values $b_1/b$ for which the polar symmetry is broken, compared to the case of a freely suspended particle. Hence, the spontaneous transition to chiral states should be easier to observe experimentally under confinement.

To summarize, we demonstrate that chirality spontaneously arises for polar active particles that attracts and repels a solute on two different portion of its surface. 
Above a critical P\'{e}clet number the polar symmetry of the solute and velocity field around the active particle breaks down and the particle spontaneously acquires an active rotation rate. Confinement between two solid parallel walls promotes the spontaneous transition to chiral states by enhancing the concentration gradient of solute around the particle. 
The symmetry breaking mechanism leading to chiral motion is very general, and hence is relevant also for active  particles which displace through different propulsion mechanisms.
Indeed, we illustrated this transition by considering a chemically-active colloid, but any other active mechanism that generates a gradient of solute would lead to a spontaneous chiralization, as long as the solute interacts differently with the two sides of the particle.
Finally, the newly discovered symmetry-breaking mechanism could be used to design microfluidic devices with controlled confinement that can selectively trap active particles in circular trajectories without the use of external fields. 

\begin{acknowledgments}
M.D.C. acknowledges funding from the European Union’s Horizon 2020 research and innovation program under the Marie Skłodowska-Curie action (GA 712754), the Severo Ochoa programme (SEV-2014-0425), the CERCA Programme/Generalitat de Catalunya and the MINECO through the Juan de la Cierva Incorporaci\'{o}n ICJ2018-035270-I. I.P. acknowledges support from MINECO Project No. PGC2018-098373-B-I00, from the DURSI Project No. 2017SGR-884, from the SNF Project No. 200021-175719 and from the EU Horizon 2020 program through 766972-FET-OPEN NANOPHLOW. G.N. acknolewdges the support of the Natural Sciences and Engineering Research Council of Canada (NSERC) Discovery Grant No. RGPIN-2017-03783. M.D.C wishes to acknowledge useful discussions with Gaetano D'Avino, Francesco Greco and Pier Luca Maffettone. The authors thank Valeria Garbin and Raymond Kapral for valuable feedbacks on an earlier version of the manuscript.
\end{acknowledgments}

\bibliography{biblio}

\end{document}